\documentstyle[preprint,aps]{revtex}
\tightenlines
\begin{document}
\draft
\title{The Fulling-Davies-Unruh Effect is Mandatory: The Proton's Testimony }

\author{George E. A. Matsas\footnote{matsas@ift.unesp.br}\\
         Instituto de F\'\i sica Te\'orica, Universidade Estadual Paulista\\ 
         Rua Pamplona 145, 01405-900, S\~ao Paulo, SP\\
	 Brazil}
\author{Daniel A. T. Vanzella\footnote{vanzella@uwm.edu}\\
        Physics Department, University of Wisconsin-Milwaukee\\ 
        1900 E Kenwood Blvd., Milwaukee, WI 53211\\
        USA}

\maketitle
\begin{abstract} 
We discuss the  {\em decay of accelerated protons}  
and illustrate how the Fulling-Davies-Unruh effect 
is indeed {\em mandatory} to maintain the consistency of 
{\em standard} Quantum Field Theory. The confidence 
level of the Fulling-Davies-Unruh effect must be 
the same as that of Quantum Field Theory itself.
\end{abstract}

\newpage
The vacuum is one of the most exquisite features of
Quantum Field Theory (QFT)~\cite{D}. Although virtual 
particles  cannot be directly detected, the explanation
of a number of effects depend on their ``existence''.
One of the most outstanding manifestations of the 
virtual particles can be found in the Casimir 
effect, which has been tested recently up to a precision 
of 1\%~\cite{MR}. The Casimir effect is a direct consequence of
the fact that the {\em metallic} 
plates disturb the {\em electromagnetic} vacuum between them. 
If metallic plates can disturb the photon vacuum, the 
curvature of the spacetime should, in general, disturb all vacua, 
since gravity couples to all the fields. The Hawking effect
is probably the most eloquent example of the particularly 
relevant r\^ole which is reserved to the concept of quantum 
vacuum in strong gravitational fields~\cite{H}. 

In 1976 Unruh~\cite{U} found that the Minkowski vacuum,
i.e., the quantum state associated with the nonexistence 
of particles with respect to inertial observers, 
corresponds to a thermal bath of particles at temperature
$T_{\rm FDU} =a/ 2 \pi $ ($\hbar = c = k =1$) to
uniformly accelerated observers with proper acceleration 
$a={\rm const}$. This has 
clarified previous results by Davies~\cite{D2}, and confirmed
Fulling's conclusion that {\em elementary particles}  are
observer dependent~\cite{F}. Roughly speaking, it can be
said that uniformly accelerated observers can see as
{\em real} those particles which inertial observers claim to
be {\em virtual}. As a result, while inertial observers would
be frozen at $0\, {\rm K}$ in the Minkowski vacuum, uniformly
accelerated observers would be heated (and possibly burnt) 
at a temperature proportional to their proper acceleration.

In spite of the fact that the Fulling-Davies-Unruh (FDU) effect can be 
rigorously derived and extended to nonlinear quantum 
fields~\cite{S} from the general Bisognano and Wichmann's 
theorem~\cite{BW}, the technicalities involved and 
probably its ``paradoxical appearance'' has kept part of 
the community quite skeptical up to now (see, e.g., Ref.~\cite{Betal}).
Many physicists, thus, have decided to ``leave the case to the experiments''. 
Notwithstanding (what would be considered) a direct manifestation 
of the FDU effect would require huge accelerations since 
$T_{\rm FDU} = [a/(2.5 \times 10^{22} {\rm cm/s}^2)] \, {\rm K}$.
Clearly, macroscopic bodies would not be able to resist such 
accelerations and, thus, from the very beginning all hopes were 
placed in elementary particle experiments~\cite{BL}.
Even so, however, paramount technical difficulties have
frustrated such endeavors although new and creative proposals
have been continuously devised~\cite{BL}-\cite{Y}
(see Ref.~\cite{R} for a comprehensive list).

Although the observation of direct manifestations of the 
FDU effect would be sympathetically received, {\em the acceptance 
that the FDU effect is mandatory to maintain (the usual) QFT consistent
should not depend on it}. Inspired by 
previous works by Unruh and Wald~\cite{UW}, Higuchi, Sudarsky 
and one of the authors have used the FDU effect 
to make sense, in the quantum realm, of the classical 
result~\cite{Roh}-\cite{B} that {\em uniformly accelerated charges 
do not radiate with respect to coaccelerated observers}. Indeed 
it was shown that {\em every Minkowski photon emitted by a 
uniformly accelerated charge (as defined in the inertial frame) 
corresponds to {\em either} the emission to {\em or} the absorption 
from the FDU thermal bath of a {\em zero-energy} Rindler photon 
(as defined in the uniformly accelerated frame)}~\cite{HMS1}. 
(Zero-energy particles 
are not detectable by physical observers, i.e., with finite proper 
acceleration.) Perhaps because the concept of {\em zero-energy} 
particles might have sounded ethereal to some, the convincing 
power of the conclusion above did not show itself to be strong enough
to reach a consensus in favor of the FDU effect. Here we call 
attention to the fact that {\em if the FDU effect does not exist, 
inertial and uniformly  accelerated observers would reach 
incompatible QFT conclusions about the stability of non-inertial protons}.

The stability of protons has been used for long time as a test for 
the standard model of elementary particles. Mounting data fix their 
lifetime  to be much longer than the present age of the universe. 
This is not so, however, if protons are accelerated rather than 
freely moving. Recently, the present authors have calculated in the context of 
{\em standard QFT} (in inertial frames) the weak-interaction decay rate for 
uniformly accelerated protons~\cite{VM}:
\begin{equation}
 p^+  \to  n^0 + e^+_M  + \nu_M
\label{pprocessinertial}
\end{equation}
and shown that in certain astrophysical situations the proton 
lifetime may be quite short.  The energy necessary to render process
(\ref{pprocessinertial}) possible is supplied by the external
accelerating agent. (We emphasize, however, that our 
final conclusion will not depend on the observation of 
the proton decay itself.) For sake of simplicity, we write the formula for the 
proton proper lifetime in {\em 1+1~spacetime 
dimensions}~\cite{MV} (rather than in 3+1~ones):~\footnote{This formula 
was derived, at the tree level, assuming a Fermi-like effective 
action through which a semi-classical current (describing the barions)
is coupled to the fermionic fields (describing the leptons). 
Eq.~(\ref{TIF}) is valid under the {\em no-recoil condition}: 
$a \ll m_p$, used to guaranty that the emitted leptons
carry small linear momentum in comparison with $m_p$, 
as measured by an inertial  observer instantaneously at rest 
with the proton.}
\begin{equation}
{\tau}^{p  \to n}_{\rm In. Observ. Calcul.} (a)
= 
\frac{2 \pi^{3/2} e^{\pi {\Delta m}/a}}{G_F^2  m_e }
\left[
G_{1\;3}^{3\;0} 
\left( \frac{m_e^2}{a^2} \left|
\begin{array}{l}
\;\;\;1\\ 
-{1}/{2}\;,\;{1}/{2}+i {\Delta m}/a\;,\;{1}/{2}-i {\Delta m}/a
\end{array}
\right.
\right) 
\right]^{-1} \; ,
\label{TIF}
\end{equation}
where 
$G_{p\; q}^{m n}$ is the Meijer function~\cite{GR},
$a$ is the proton proper acceleration,
$ \Delta m \equiv m_n - m_{p }$ and we have 
assumed $m_\nu = 0$. (Here $m_p$, $m_n$, $m_e$ and $m_\nu$
are the rest masses of the proton, neutron, electron and neutrino,
respectively.) 
The value of the effective Fermi constant $G_F=9.9 \times 10^{-13}$
is determined from phenomenology. (Because in four dimensions inertial 
neutrons decay in 887 s, we have chosen this value for the neutron 
lifetime in two dimensions as well.) 

Since a uniformly accelerated proton can be confined in a 
Rindler wedge which is a globally hyperbolic spacetime possessing
a global timelike isometry, the associated uniformly accelerated
observers (Rindler observers) must be able to analyze this 
phenomenon and reobtain the same (scalar) value for the proton lifetime
(\ref{TIF}) 
obtained with standard QFT. Notwithstanding, because of energy
conservation, Rindler observers would simply claim that protons are
precluded from decaying into a neutron through 
\begin{equation}
 p^+  \to  n^0 + e^+_R  + \nu_R \; .
\label{pprocessinertial'}
\end{equation}
Hence an extra ingredient must be added, otherwise inertial and Rindler 
observers would conclude precisely the opposite about the stability of
uniformly accelerated protons. This extra ingredient is the FDU thermal 
bath. According to the Rindler observers, the Minkowski vacuum is ``seen''
as a thermal bath with which the proton may interact. As a consequence,
new channels are opened:
\begin{equation}
p^+ + e^-_R \to  n^0 + \nu_R
\label{pprocessaccelerated1}
\end{equation}
\begin{equation}
 p^+ + \bar \nu_R  \to  n^0 + e^+_R 
\label{pprocessaccelerated2}
\end{equation}
\begin{equation}
p^+  + \bar\nu_R + e^-_R \to  n^0  \; ,
\label{pprocessaccelerated3}
\end{equation}
i.e., from the point of view of the Rindler observers, the proton 
should be transformed 
into a neutron through the {\em absorption} of a Rindler electron and/or 
anti-neutrino from the surrounding thermal bath providing the 
necessary energy to allow the process to occur. Eventually
any energy in excess can be disposed by the emission of a neutrino 
or a positron (depending on the case).  
Indeed, by performing an independent QFT calculation in the uniformly 
accelerated frame, 
we have obtained the following proper lifetime for the proton, after combining 
(incoherently) processes 
(\ref{pprocessaccelerated1})-(\ref{pprocessaccelerated3})
in the presence of the FDU thermal bath:
\begin{equation}
{\tau}^{p  \to n}_{\rm Rin. Observ. Calcul.}
=
\frac{\pi^2 a e^{\pi \Delta m/a}}{G_F^2 m_e}
\left[
\int_{-\infty}^{+\infty} d {\omega_R}
\frac{ 
      K_{i {\omega_R}/a + 1/2} ( m_e/a)
      K_{i {\omega_R}/a - 1/2} ( m_e/a)   
     }
     {\cosh[\pi ({\omega_R} -{\Delta m})/a ] }
\right]^{-1}
 \, .
\label{TAF}
\end{equation}
Although Eqs.~(\ref{TIF}) and (\ref{TAF}) appear to be quite different, we
have shown numerically that they coincide up to basically the 
machine-precision limit~\cite{VM2}:
$$
\Delta^{-1} \int_\Delta dx 
[({\tau}^{p  \to n}_{\rm Rin. Observ. Calcul.}
-
  {\tau}^{p  \to n}_{\rm In. Observ. Calcul.})/
  {\tau}^{p  \to n}_{\rm Rin. Observ. Calcul.}]^2
\sim 10^{-16},
$$
where $x \equiv log_{10} (a/1 MeV)$. 

The FDU effect is not only very important for its own right but is also useful 
as a guide in the investigation of some unresolved questions of the
Hawking effect as, e.g., the r\^ole played by the 
transplackian frequencies (see Ref.~\cite{Wrev} and references 
therein). The FDU effect must be seen as being as necessary
to QFT as the non-inertial forces  (centrifugal and Coriolis ones) 
are to Mechanics since both are required to maintain successfully
tested theories consistent when analyzed in non-inertial frames.
The description of the decay of non-inertial protons in the uniformly
accelerated frame {\em in the absence of the FDU thermal bath} is 
the {\em challenge} which the present Essay poses for those who 
are still doubtful about this effect.

\begin{flushleft}
{\bf{\large Acknowledgements}}
\end{flushleft}

G.M. has long been indebted to A. Higuchi, D. Sudarsky and R. Wald 
for many enlightening discussions on QFT in curved spacetimes.
The authors would like to acknowledge particularly A. Higuchi
for discussions in the early stages of this research. We are also thankful
to G. Howells for improvements in the manuscript. G.M. and D.V. were 
partially supported by the following Brazilian agencies: Conselho Nacional 
de Desenvolvimento Cient\'\i fico e Tecnol\'ogico and Funda\c c\~ao de 
Amparo \`a Pesquisa do Estado de S\~ao Paulo, respectively. 
This work was also supported in part by the US National Science Foundation 
under Grant No. PHY-0071044.

\end{document}